\documentclass[preprint, showpacs,amsmath,amssymb]{revtex4}
\usepackage{graphicx} %include figure files
\usepackage{dcolumn}%alingn table columns on decimal point
\usepackage{bm}%bold math
\begin{document}
\title{Influence of subband mixing, due to spin-orbit interaction,\\  
on the  transmission through  periodically modulated waveguides}
\author{ X. F. Wang$^\dagger$ and 
P. Vasilopoulos$^\star$ \\
\ \\}
\address{ Concordia University Department of Physics\\
 1455 de Maisonneuve Ouest \\ Montr\'{e}al, Qu\'{e}bec,   H3G  1M8,  Canada\\
 \ \\  }
%\maketitle
\date{\today} 
\begin{abstract}
Ballistic spin transport, through periodically stubbed waveguides,  
is studied in the presence of a weak spin-orbit interaction (SOI) and  the resulting subband mixing.
By an appropriate choice of the waveguide length and of the stub parameters
injected spin-polarized electrons can be blocked completely and the 
transmission shows a periodic and nearly square-wave pattern with wide gaps when only one mode is allowed to propagate
in the waveguide. Relative to the case when  subband mixing is neglected, the  transmission  changes drastically as a
function of the incident electron energy or of the stub height, as it exhibits new peaks or dips, but remains robust as
a function of the stubs' degree of asymmetry. Varying the strength of the SOI parameter changes the relative
contribution to the total transmission or conductance of the spin-up and spin-down states. The structure considered is a
reasonable candidate for establishing a spin transistor.
%****
\end{abstract}
\maketitle  
\clearpage
\section{INTRODUCTION}

Recently research in spin-related effects, such as  spin injection into devices,  spin-polarized transport, etc.,
has been intensified. Part of the reason is that the possibility exists to use the electron's spin for quantum
computations \cite{kik}. The  basic principle of a spin transistor was  formulated in Ref. \cite{dat} for a waveguide 
in the presence of  the spin-orbit interaction (SOI) or Rashba coupling  \cite{ras} and was recently studied for a
 simple semiconductor waveguide using a tight-binding model \cite{mir} or periodically modulated waveguides \cite{wan}. 
Various  spin-filtering \cite{gove} or spin-valve \cite{mire} effects have  been studied and several designs have been
proposed to spin-polarize electronic  currents in nanostructures \cite {kis, gove}
among other studies of SOI effects on the band structure and transport of similar systems \cite{moro, gove1}. 
%****
In this respect several efforts 
have been made using ferromagnet-semiconductor interfaces to produce spin-polarized electrons, but this method must face
the mismatch of physical parameters between these two quite different materials \cite{zhu}.  Another idea is to  employ
diluted magnetic semiconductors 
(DMS), which can match well with other extensively used semiconductors
like AlGaAs, and has created a lot of interest in DMS \cite {ohn,luc}.

Spin degeneracy  in semiconductors  results from inversion
symmetry, in space and time, of the considered system.  By introducing
a spatial inversion asymmetry, one can realize spin splitting for
carriers of finite momentum without applying any external magnetic
field.  This so-called Rashba effect or spin-orbit interaction (SOI) \cite{ras,win}
has been confirmed experimentally in various semiconductor
structures \cite{das}.  In semiconductor
heterostructures this spatial inversion asymmetry can be easily
obtained by either built-in and external electric fields or by the
position-dependent band edges.  It is found that in many cases, 
especially in narrow gap semiconductor structures, the corresponding 
SOI is a linear function of the electronic momentum 
$\mathbf{k}$\ expressed as the Rashba term $\overrightarrow{\sigma} \cdot (\mathbf{k\times E)}$ in the electron
Hamiltonian, where $\overrightarrow{\sigma }$ is the Pauli spin matrix and $\mathbf{E}$ 
the local electric field.  Thus, a local electric field $\mathbf{E}$ acts on the 
electronic spin like a local magnetic field perpendicular to the 
directions of $\mathbf{E}$  and of the electron momentum.  The 
Rashba parameter is proportional to the average value of $\mathbf{E}$  and can
be well controlled by a top (back) gate over (below) the device \cite{nit}.

Ballistic {\it spinless} electronic transport has been studied extensively in systems of reduced dimensionality 
\cite{sols,tak1} but until recently it was not  known how to effectively control the 
spin-polarized flux in the relevant  systems. In  previous work \cite{wan} we 
showed  how spin-polarized transport can be realized and 
controlled in stubbed waveguides mostly when only spin-up or spin-down 
electrons are injected and only one mode  propagates in the waveguide. Our treatment, which showed how a
{\it square-wave} and spin-dependent transmission could be realized, 
relied on the weakness of the SOI and the neglect of subband mixing due to this interaction. 
This put some constrains on the ranges of various parameters, namely  the width of the waveguide,
 the height of the attached stubs, and the energy of the incident electrons.
In the present paper we build upon this work and study in detail
the effects of subband mixing which, to our knowledge, have been dealt with only partly in Ref. \cite{mir}
for waveguides without stubs. In doing so we relax substantially the constrains mentioned above.  
Again  our aim is to investigate in detail the conditions for the realization of  a spin 
transistor in periodically stubbed semiconductor waveguides in the  presence of SOI. 
As will be shown, we find new results (peaks or dips) in the transmission as a function of the incident electron energy
or of the stub height,  but its square-wave pattern, as a function of the stubs' degree of asymmetry, remains robust. 
  
In Sec.  II we present the formalism and contrast the results for the eigenvalues and eigenvectors with those obtained
without subband mixing. In Sec.  III  we formulate the transmission problem and in Sec. IV we present  numerical
results. Concluding remarks follow in Sec.  V.

\section{FORMALISM}

\begin{figure}[tpb]
%\vspace{-2.5cm}
\includegraphics*[width=75mm,height=90mm]{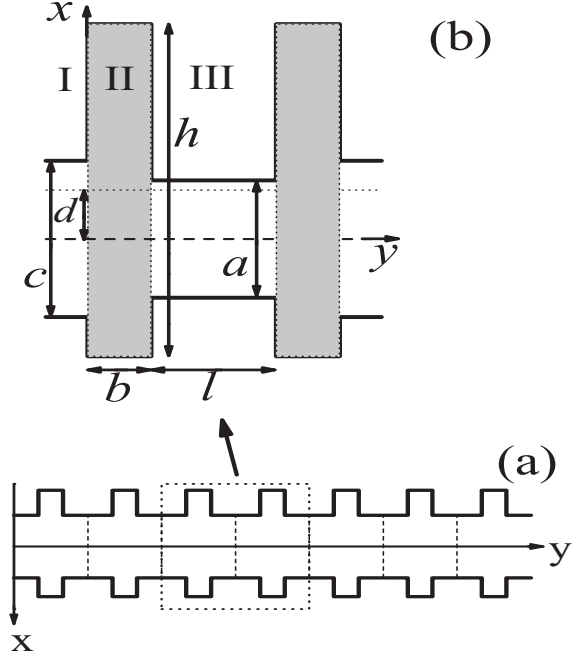}
%\vspace{-1.5cm}
\caption{
A stubbed waveguide along the $y$ direction (a) with a two-stub section detailed in (b).
The width is $c$ in region I and $a$ in region III.
The height of the stubs  is $h$, their length $b$, and the length between them $l$. 
The asymmetry parameter $d$ is the distance between the centerline of the waveguide and stubs.}
\label{fig1}
\end{figure}

When a typical two-dimensional (2D) electronic system, such as an InGaAs/InAlAs
quantum well, is confined, e.g., by a potential $V(x)$ along the $x$ direction,
we have a quasi-one-dimensional (Q1D) electronic system such as the stubbed waveguide shown
in Fig. \ref{fig1}. The one-electron Hamiltonian,
including the Rashba SOI term, reads

\begin{equation}
H=\frac{\vec{p}^{2}}{2m^{\ast }}+\frac{\alpha }{\hbar 
}(\vec{\sigma}
\times \vec{p})_{z} +V(x)
=\left[
\begin{array}{cc}
-\lambda\vec{\nabla}^{2}+V(x) & \alpha\nabla^-\\
-\alpha\nabla^+ & -\lambda\vec{\nabla}^{2}+V(x)
\end{array}
\right],
\end{equation}
 where $\lambda=\hbar ^2/2m^*$, $\vec{\nabla}^{2}=\partial ^{2}/\partial 
 x^{2}+\partial ^{2}/\partial y^{2}$, and
$\nabla_{\pm}= \partial/\partial x\pm i\partial/\partial y$.
The parameter $\alpha $
measures the \noindent strength of the SOI
and is proportional to the interface electric field; $\vec{\sigma}=(\sigma
_{x},\sigma _{y},\sigma _{z})$ denotes the spin Pauli matrices, and
${\vec{p}}$ is the momentum operator. The wave function can be expressed in the form
\begin{equation}
\Psi_{k_{y}}({\bf r})=e^{ik_{y}y}\sum_{n\sigma }\phi_n(x)C_n^{\sigma }|\sigma\rangle
=e^{ik_{y}y}\sum_n\phi_n(x)\left(
\begin{array}{c}
C_n^{+}\\ 
C_n^{-}
\end{array}
\right),
\end{equation}
with $|\sigma \rangle ={\tiny\left(
\array{c}
1\\0
\endarray
\right)}$ for spin up (+) and ${\tiny \left(
\array{c}
0\\1
\endarray
\right)}$ for spin down (-). 
$\phi(x)$ is the eigenfunction of the 1D Hamiltonian
$h(x)=-\lambda\nabla_x^{2}+V(x)$ with an assumed square-well confining potential $V(x)$.

We insert this eigenfunction in the equation $H\Psi =E\Psi $, multiply both sides
by $\phi _{m}(x)$, and integrate
over $x$. With $\int dx\phi _{m}(x)\phi _{n}(x)=\delta _{mn}$
and $J_{mn}=\int dx\phi _{m}(x)\phi _{n}^{\prime}(x)$ we obtain
\begin{equation}
\left[
\begin{array}{cc}
E_{m}^0 -E & \alpha k_{y}\\
\alpha k_{y} & E_{m}^0-E
\end{array}
\right]
\left(
\begin{array}{c}
C_{m}^{+}\\
C_{m}^{-}
\end{array}
\right)
+\alpha\sum_n
J_{mn}
\left[
\begin{array}{cc}
0 & 1 \\
-1 & 0 
\end{array}
\right]
\left(
\begin{array}{c}
C_{n}^{+} \\
%\\
C_{n}^{-}
\end{array}\right)
=0,
\label{crosse}
\end{equation}
where $E_{m}^0=E_{m}+\lambda k_{y}^{2}$; the index $m$ labels the discrete subbands resulting from the confinement along
the $x$ axis.

As shown elsewhere \cite{wan}, if we neglect mixing between the 
subbands by assuming $J_{mn}\approx 0$, we can easily solve Eq. 
(3).  This procedure is valid for $ | \alpha J_{mn}|\ll 
|E_{m}-E_{n}|$. The resulting eigenvalues  read 
\begin{equation}
E^{\pm }(k_{y})=
E_{m}+\hbar ^{2}k_{y}^{2}/2m^{\ast }\pm \alpha 
k_{y}.
\label{disp}
\end{equation}
The eigenvectors corresponding to $E^{+},E^{-}$ satisfy $C_{m}^{+}=\pm
C_{m}^{-}$. Accordingly, the spin eigenfunctions can be taken as
\begin{equation} |\pm \rangle =
{\tiny\left(
\array{c}
1\\
\pm 1 \endarray \right)/\sqrt{2}}.
\label{eig}
\end{equation}
An important aspect in this case is that the difference in wave 
vectors $k_{y}^{+}$ and $k_{y}^{-}$, resulting from $ 
E^{+}=E^{-}=E$, is constant: it reads
\begin{equation}
k_{y}^{-}-k_{y}^{+}=2m^{\ast }\alpha /\hbar ^{2}.
\label{delt}
\end{equation}

To go beyond this limiting case, described by $J_{mn}\approx 0$, 
and still have a tractable problem, we neglect  all $J_{mn}$ 
terms except $J_{21}$ and $J_{12}=-J_{21}=8/3w=\delta$, where $w$ is the width 
of the waveguide along $x$. That is, we  include   mixing only 
between the first and  second subband. Then the secular equation for 
these two lowest subbands reads
%
%***
\begin{equation}
\left[
\begin{array}{cccc}
E_{1}^0 -E & \alpha k_{y}&0&\alpha \delta\\
\alpha k_{y} & E_{1}^0 -E &-\alpha \delta &0\\
0&-\alpha \delta &E_{2}^0 -E & \alpha k_{y}\\
\alpha \delta &0&\alpha k_{y} & E_{2}^0 -E \\
\end{array}
\right]
\left(
\begin{array}{c}
C_{1}^{+} \\
C_{1}^{-}\\
C_{2}^{+} \\
C_{2}^{-}
\end{array}
\right)=0.
\end{equation}
The resulting eigenvalues ($\varepsilon_n^{\sigma}$)
and eigenvectors ($\Psi_n^{\sigma}$) are
\begin{equation}
\left\{
\begin{array}{ll}
\varepsilon_1^+=
(E^0_1+E^0_2-\Delta E_-)/2, &
\ \ \ \ \ \ \ \ \Psi_1^+=\frac{1}{C }
\left(\begin{array}{c}
\phi_1+r_B\phi_2\\%\frac{
\phi_1-r_B \phi_2
\end{array}
\right),\\
% 1
\varepsilon_1^-=
(E^0_1+E^0_2-\Delta E_+)/2, &

\ \ \ \ \ \ \ \ \Psi_1^-=\frac{1}{D}
\left(\begin{array}{c}
-\phi_1+r_A\phi_2\\ 
\phi_1+r_A\phi_2 
\end{array}
\right),\\
% 2
\varepsilon_2^+=
(E^0_1+E^0_2+\Delta E_+)/2, &
\ \ \ \ \ \ \ \ \Psi_2^+=\frac{1}{D}
\left(\begin{array}{c}
\phi_2+r_A\phi_1\\%\frac{2\alpha\delta}{A}\phi_1\\
\phi_2-r_A\phi_1 
\end{array}
\right),\\
% 3
\varepsilon_2^-=
(E^0_1+E^0_2+\Delta E_-)/2, &
\ \ \ \ \ \ \ \ \Psi_2^-=\frac{1}{C}
\left(\begin{array}{c}
-\phi_2+r_B\phi_1\\
\phi_2+r_B\phi_1
\end{array}
\right),\\
% 4
\varepsilon_n^+=E^0_n+\alpha k_y, \ \ \ \ \ \  n > 2; &
\ \ \ \ \ \ \ \ \Psi_n^+=\frac{1}{\sqrt{2}}
\left(\begin{array}{c}
\phi_n\\
\phi_n
\end{array}
\right),\\
% 5
\varepsilon_n^-=E^0_n-\alpha k_y, \ \ \ \ \ \  n > 2; &
\ \ \ \ \ \ \ \ \Psi_n^-=\frac{1}{\sqrt{2}}
\left(
\begin{array}{c}
-\phi_n\\
\phi_n
\end{array}
\right).\\
% 6
\end{array}\right.
\end{equation}
Here $\Delta E_\pm=[(\Delta E_{12}\pm 2\alpha 
k_y)^2+4\alpha^2\delta^2]^{1/2}$, $\Delta E_{12}=E_2^0-E_1^0$, 
$A=(\Delta E_{12}+2\alpha k_y)+\Delta E_+$, $B=(\Delta 
E_{12}-2\alpha k_y)+\Delta E_-$, $r_A=2\alpha\delta/A$, 
$r_B=2\alpha\delta/B$, $D^2=2+2r_A^2$, and $C^2=2+2r_B^2$. Notice 
that the first four {\it two-row} eigenvectors are linear 
combinations of the {\it four-row} ones corresponding to Eq. (6). We further notice that the eigenvalues
and eigenvectors given above reduce to those given, respectively, by Eqs. (4) and (5) if we set $\delta=0$.

\begin{figure}[tpb]
\vspace{-1.5cm}
\includegraphics*[width=80mm,height=70mm]{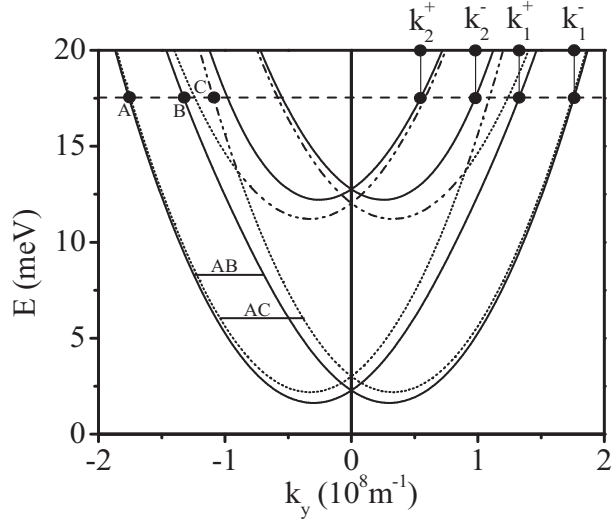}
\vspace{-0.5cm}
\caption{Energy dispersion of the two lowest subbands, $E_1$ and $E_2$, in a InGaAs 
waveguide  500\AA \ wide. The solid curves  include subband mixing, 
induced by  the SOI, the dotted and dash-dotted ones do not. The
intersections of the dispersion curves with the dashed line, showing the constant energy
of the incident electrons, defines the $\pm$ components of the wave vectors.} 
\label{fig2}
\end{figure}

The  dispersion relation given by Eq. (8) for the lowest two 
subbands  is shown in Fig. \ref{fig2} by the solid curves. For an 
electron of energy $E$ in branches $\varepsilon_1^+$, 
$\varepsilon_1^-$, $\varepsilon_2^+$, and $\varepsilon_2^-$, the 
corresponding wave vectors along $y$ are $k_1^+$ ($-k_1^-$), 
$k_1^-$ ($-k_1^+$), $k_2^+$ ($-k_2^-$), and $k_2^-$ ($-k_2^+$), 
respectively. When an electron with a positive wave vector $k_y$ 
has an energy much higher than $E_c=5E_1/2+9\hbar^2 
E_1^2/(8m^*\alpha^2)$, the value at which anticrossing due to the 
SOI occurs between the $\varepsilon_1^+$ and the 
$\varepsilon_2^-$ branches, its spin is up along the $x$ direction 
($|\sigma\rangle =\tiny\left(\begin{array}{c} 1 \\1 \end{array} 
\right)$) when it is in the two higher spin branches 
($\varepsilon_2^+$, and $\varepsilon_2^-$) and  down 
($|\sigma\rangle =\tiny\left(\begin{array}{c}1 
\\-1\end{array}\right)$) when it is in the other two branches. 
When its energy is lower than $E_c$, its spin is up   in the 
"+" branches ($\varepsilon_1^+$, and $\varepsilon_2^+$) and 
down in the  "-" branches. Electrons of the same energy and the 
same but opposite momentum have always opposite spin orientation. 
Similar observations were made in Ref. \cite{gove1} for a quantum wire with parabolic
confinement.
One interesting case is that an electron of positive momentum 
always has its spin pointing down and vice versa in the bag of 
each branch  $\varepsilon_n^{\sigma}$ with energy between the 
bottom of the branch and zero-momentum energy $\varepsilon^0_n$.
Another noteworthy feature in Fig. 2 is that mode mixing makes the wave
vector differences $k_1^+-k_1^-$ and  $k_2^+-k_2^-$ depend slightly on the energy. The length of  the horizontal segment
AC between the dotted curves, given by Eq. (6), is constant and independent of the energy whereas that of the segment AB
between the solid curves is not and depends on the energy; AB satisfies Eq. (6) only approximately.

\section{Formulation of the transmission problem}

Let us consider the transmission process when an electron of
energy $E$ is incident from the left to a stubbed waveguide shown in 
Fig. \ref{fig1}. The electron wave function is decomposed into "+"
and "-" branches in  all  regions  in Fig.
\ref{fig1}.  In each region we have $\phi _{n}(x)=\sin (n\pi (x+w/2)/w)$, $0\leq x\leq w$, 
where $w$ is the width along the $x$ direction.  
Including spin and referring to Fig.  \ref{fig2}, we can write the
eigenfunction of energy $E$ in region I as
\begin{eqnarray}
\phi _{1} &=&\sum_{m}[
c_{m}\Psi_{mc}^+(\eta^+_m)e^{i\eta^+_{m}y}+
\bar{c}_{m}\Psi_{mc}^+(-\eta^-_m)e^{-i\eta^-_m y}\nonumber\\
&+&d_{m}\Psi_{mc}^-(\eta^-_m)e^{i\eta^-_{m}y}+
\bar{d}_{m}\Psi_{mc}^-(-\eta^+_m)e^{-i\eta^+_m y}],
\end{eqnarray}
in region III as
\begin{eqnarray}
\phi _{2} &=&\sum_{m}[
a_{m}\Psi_{ma}^+(\beta^+_m)e^{i\beta^+_{m}(y-b)}+
\bar{a}_{m}\Psi_{ma}^+(-\beta^-_m)e^{-i\beta^-_m (y-b)}\nonumber\\
&+&b_{m}\Psi_{ma}^-(\beta^-_m)e^{i\beta^-_{m}(y-b)}+
\bar{b}_{m}\Psi_{ma}^-(-\beta^+_m)e^{-i\beta^+_m (y-b)}],
\end{eqnarray}
and in the stub region II as
\begin{eqnarray}
\phi _{s} &=&\sum_{m}[
u_{m}\Psi_{mh}^+(\gamma^+_m)e^{i\gamma^+_{m}y}+
\bar{u}_{m}\Psi_{mh}^+(-\gamma^-_m)e^{-i\gamma^-_m y}\nonumber\\
&+&v_{m}\Psi_{mh}^-(\gamma^-_m)e^{i\gamma^-_{m}y}+
\bar{v}_{m}\Psi_{mh}^-(-\gamma^+_m)e^{-i\gamma^+_m y}].
\end{eqnarray}
Here $\eta_m^\pm$, $\beta^\pm_m$, and $\gamma^\pm_m$ are the wavevectors $k^{\pm}_m$ in regions I, III, and II,
respectively.  In this paper we study the case where the electron energy is low enough so 
that at most  two modes propagate in the waveguide segments
(region I and III) though more modes are considered in the stubs (region II). We proceed as follows.

We  match the wave functions of different
regions at $y=0$ and $y=b$: we multiply by
$\Psi_{1h}^+(\gamma^+_1)$, $\Psi_{1h}^-(\gamma^-_1)$,
$\Psi_{2h}^+(\gamma^+_2)$, and $\Psi_{2h}^-(\gamma^-_2)$,
respectively, the equations $\Psi_s(y=0)=\Psi_1(y=0)$ and   
 $\Psi_s(y=b)=\Psi_2(y=b)$. Then integrating over $x$ we
obtain eight linear equations for the eight coefficients of the
wave functions of the  two lowest coupled subbands
denoted by the matrices
$\hat{U}_{12}^T=(u_1,\bar{u}_1,v_1,\bar{v}_1,u_2,\bar{u}_2,v_2,\bar{v}_2)$,
in region II,  
$\hat{L}_{12}^T=(c_1,\bar{c}_1,d_1,\bar{d}_1,c_2,\bar{c}_2,d_2,\bar{d}_2)$
in region I, 
and by
$\hat{R}_{12}^T=(a_1,\bar{a}_1,b_1,\bar{b}_1,a_2,\bar{a}_2,b_2,\bar{b}_2)$
in region III, where $T$ denotes the transfer  matrix.
This gives
\begin{equation}
\hat{M}_{12}\hat{U}_{12}=\hat{P}_{12}\hat{L}_{12}+\hat{Q}_{12}\hat{R}_{12}.
\end{equation}
The coefficients corresponding to $n>2$, $\hat{U}_{n}^T=(u_n,\bar{u}_n,v_n,\bar{v}_n)$,
can be found in a similar way by multiplying by
$\Psi_{nh}^+(\gamma^+_n)$ and $\Psi_{nh}^-(\gamma^-_n)$ before integrating over $x$; the result is
\begin{equation}
\hat{M}_{n}\hat{U}_{n}=\hat{P}_{n}\hat{L}_n+\hat{Q}_{n}\hat{R}_{12}.
\end{equation}
The matrices 
$\hat{M}_{12}, \hat{P}_{12}, \hat{Q}_{12}, \hat{M}_{n}, \hat{P}_{n},\hat{Q}_{n}$
as well as the matrices
$\hat{N}_{12},  \hat{N}_n, \hat{U}_n, \hat{\eta}_{12},\hat{\beta}_{12}$, appearing in Eqs. (14) and (15), are specified
in the appendix.

We now match  the derivatives of the wave functions at $y=0$ and
$y=b$ and multiply  by $\Psi_{1c}^+(\eta^+_1)$,
$\Psi_{1c}^-(\eta^-_1)$, $\Psi_{2c}^+(\eta^+_2)$,
$\Psi_{2c}^-(\eta^-_2)$  the equation
$d\Psi_s/dy|_{y=0}=d\Psi_1/dy|_{y=0}$ and by
 $\Psi_{1a}^+(\beta^+_1)$, $\Psi_{1a}^-(\beta^-_1)$,
$\Psi_{2a}^+(\beta^+_2)$, $\Psi_{2a}^-(\beta^-_2)$ the equation
$d\Psi_s/dy|_{y=b}=d\Psi_1/dy|_{y=b}$. Then we integrate over $x$
and obtain 
\begin{equation}
\hat{N}_{12}\hat{U}_{12}+\sum_n\hat{N}_n\hat{U}_n=\hat{\eta}_{12}\hat{L}_{12}+\hat{\beta}_{12}\hat{R}_{12}.
\end{equation} 
Now the relation between the coefficients of the wave function
to the left of the waveguide (region I) and to its right (region
III) is established as
\begin{equation}
\hat{L}_{12}=(\hat{N}_{12}\hat{M}^{-1}_{12}\hat{P}_{12}+
\sum_{n>2}\hat{N}_n\hat{M}^{-1}_{n}\hat{P}_n-\hat{\eta}_{12})^{-1}
(-\hat{N}_{12}\hat{M}^{-1}_{12}\hat{Q}_{12}-
\sum_{n>2}\hat{N}_n\hat{M}^{-1}_{n}\hat{Q}_n+\hat{\beta}_{12})
\hat{R}_{12}=\hat{T}_1\hat{R}_{12}.
\end{equation}
If there are more than one unit in the device, we  denote the
transfer matrix of the $i$-th stub as $\hat{T}_i$,
 that of the $i$-th waveguide segment as $\hat{P}_i$, and obtain the total transfer matrix as
\begin{equation}
\hat{T}=\prod_i \hat{T}_i \hat{P}_i.
\end{equation}

Assuming we input electrons from the left of the device and
measure the transmission at its right, the reflection coefficient  
at its right should be zero. For $E\geq\varepsilon^0_1=\{E_1+E_2-[(\Delta E_{12})^2+4\alpha^2\delta^2]^{1/2}\}/2$, with
$\varepsilon^0_1$ the first-subband's zero-momentum energy,  we have
$\bar{a}_1=\bar{b}_1=0$, and for $E\geq\varepsilon^0_2=\{E_1+E_2+[(\Delta
E_{12})^2+4\alpha^2\delta^2]^{1/2}\}/2$, with $\varepsilon^0_2$  the second-subband's
zero-momentum energy, we have $\bar{a}_2=\bar{b}_2=0$. The transmission
matrix $\hat{M}_t$ and the reflection matrix $\hat{M}_r$ are given by
\begin{equation}
\hat{M}_t^{-1}=
\left[\begin{array}{cccc}
T_{11}&T_{13}&T_{15}&T_{17}\\
T_{31}&T_{33}&T_{35}&T_{37}\\
T_{51}&T_{53}&T_{55}&T_{57}\\
T_{71}&T_{73}&T_{75}&T_{77}\\
\end{array}
\right],
\end{equation}
 and 
\begin{equation}
\hat{M}_r\hat{M}_t^{-1}=
\left[\begin{array}{cccc}
T_{21}&T_{23}&T_{25}&T_{27}\\
T_{41}&T_{43}&T_{45}&T_{47}\\
T_{61}&T_{63}&T_{65}&T_{67}\\
T_{81}&T_{83}&T_{85}&T_{87}\\
\end{array}
\right],
\end{equation}
where $T_{ij}$ is the element of the transfer matrix $\hat{T}$.
The transmission process is then embodied in the
matrix $\hat{M}_t$:
\begin{equation}
\left(
\begin{array}{c}
a_1\\b_1\\a_2\\b_2\\
\end{array}
\right)
=\hat{M}_t
\left(
\begin{array}{c}
c_1\\d_1\\c_2\\d_2.
\end{array}
\right)
\end{equation}

\section{results and discussions}

In our previous work \cite{wan} we obtained a {\it square-wave} spin
transmission as a function of $h$ and $d$ when one mode
propagates in
the waveguide and the subband mixing due to the SOI is negligible. However,
when a gate voltage is applied to the stubs to increase their
height $h$, the second subband approaches  the first one and
the mixing between them  becomes stronger.
If not otherwise specified, we consider only spin-up incident electrons and the
following parameters: width $a=c=500$ \AA, stub height $h=1600$\AA, stub length $b=660$\AA, waveguide segment length
$l=1050$\AA,
asymmetry parameter $d=0$, electron energy $E=4$meV, and   
$\alpha=1.6\times 10^{-11}eVm$. 
The conductance $G$ at zero temperature can be expressed in terms of the transmission $T$ as
$G=(e^2/h)T$. 
%????

\begin{figure}[tpb]
\vspace{-1cm}
\includegraphics*[width=70mm,height=70mm]{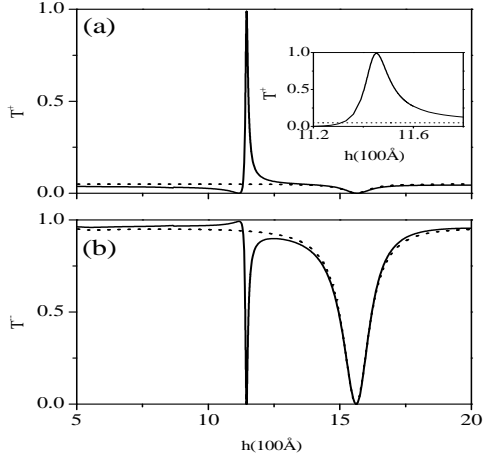}
\vspace{-1cm}
\caption{Transmission $T$ as a function of the stub height
$h$ for \underline{one} stub. Panel  (a) is for {\it spin up}
and panel (b) for {\it spin down}. The  solid curves include subband mixing,
the dotted ones do not. Notice the absence
of the first peak in (a) and the first dip in (b) when mixing is neglected.
The inset in (a) is a detailed view of the first peak.}
\label{fig3}
\end{figure}

In Fig. \ref{fig3}, we show the spin-up (a) and spin-down transmission (b) through one unit with one symmetric stub as
a function of the stub height $h$ when only spin-up electrons are incident.
The solid curves denote the results with mixing and the dotted ones without it. The only
influence of the mixing in (a) is that a nearly zero transmission is followed by a transmission peak
in the range   $1100\AA \leq h \leq \ 1200$\AA. The form of the peak is
shown in detail in the inset. In  contrast, in the same range of $h$ the
spin-down transmission in (b) shows a dip instead of a peak.
This happens when $h$ has such a value that the first and the second subbands in the stubs are coupled with each other
by the mixing term $J_{mn}$ and both of them are coupled well with the waveguide mode through the interface connecting
the stub and waveguide. The numerical result shows that the phase of the output electrons
 
\noindent
is changed by the shift of $h$ when this transmission
oscillation happens, which has not been observed when mixing is neglected \cite{wan}.
Correspondingly, the phase difference between the "+" and "-" branches
and the spin orientation of the output electrons is changed by $h$ but the total transmission
is kept constant. Here we see that one  important effect resulting from SOI mixing is that
the precession of the electronic spins depends not only on the length of the waveguide but also on its shape and
width. If we change the  parameter $\alpha$ to shift the anticrossing energy $E_c$, the position of the oscillation does
not shift but its amplitude can change. For the parameters used here
the electron energy is close to $E_c$.

\begin{figure}[tpb]
\vspace{-1cm}
\includegraphics*[width=70mm,height=70mm]{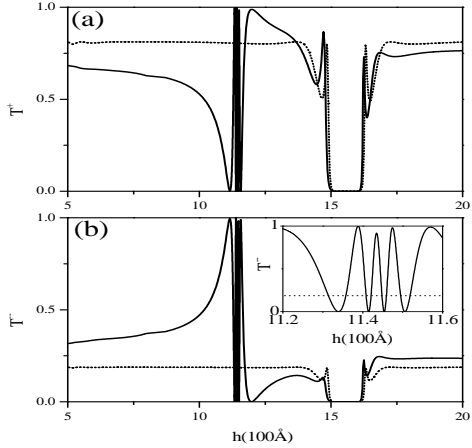}
\vspace{-1cm}
\caption{As in Fig. 3 but for \underline{five} stubs.
Notice the wide gap for both spin states starting at about
$h$=1500\AA. Again the results without mixing (dotted curves) miss the prominent structure near $h$=1120\AA.
The inset in (b) is a detailed view of the region $1120$\ \AA$\ \leq h \leq 1160$\ \AA.}
\label{fig4}
\end{figure}
Next we evaluate the transmission of a structure composed of five units identical to the one above.
We obtain a square-wave 
pattern of the transmission if we neglect mixing, as shown by the
dotted curves in Fig. \ref{fig4}. Surprisingly, the subband mixing 
does not change the square-wave form of the  
transmission gap. Nevertheless, both spin-up and spin-down transmissions
shift here and there and five oscillations appear for
$h=1120$\AA. The inset in Fig. \ref{fig4} (b) shows in detail these oscillations.
It is worth noting that here the transmission is much more sensitive to the variation of $h$ than that in the previous
one-unit case and its oscillations may be weakened or rounded off by  lateral fluctuations of $h$ which are expected to
occur in  real nanostructures.
%***

\begin{figure}[tpb]
%\vspace{-1.7cm}
\includegraphics*[width=70mm,height=70mm]{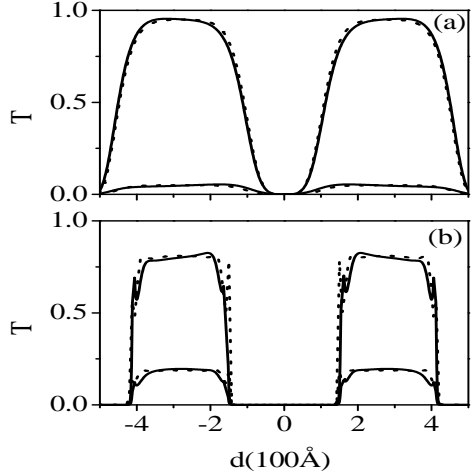}
\vspace{-0.4cm}
\caption{Transmission $T$ as a function of the asymmetry
parameter  $d$ for {\it one} stub in (a) and {\it five} stubs
in (b).
The lower (upper) curves in (a) and the upper (lower) ones in (b) are for spin up (down).
The  solid curves include subband mixing, the
dotted ones do not. As shown, the  differences between
the solid and dotted curves are minimal.}
\label{fig5}
\end{figure}

\begin{figure}[tpb]
\vspace{-0.5cm}
\includegraphics*[width=90mm,height=70mm]{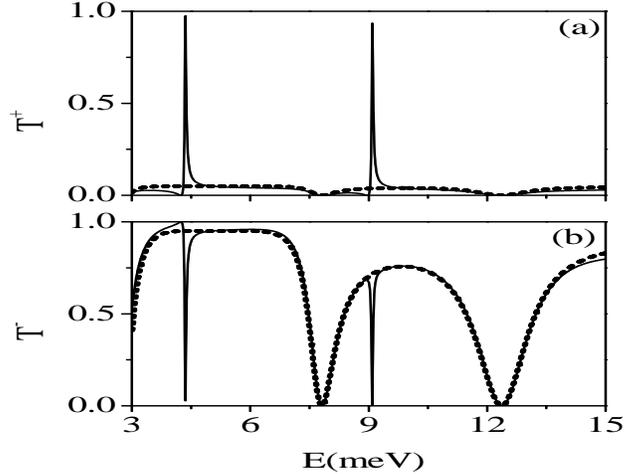}
\vspace{-0.5cm}
\caption{Transmission $T$ versus
incident electron energy for one stub. Panel (a) is for {\it spin up}
and panel (b) for {\it spin down}. The  solid curves include subband mixing, the dotted ones do not.}
\label{fig6}
\end{figure}
\begin{figure}[tpb]
\vspace{-3cm}
\includegraphics*[width=70mm,height=70mm]{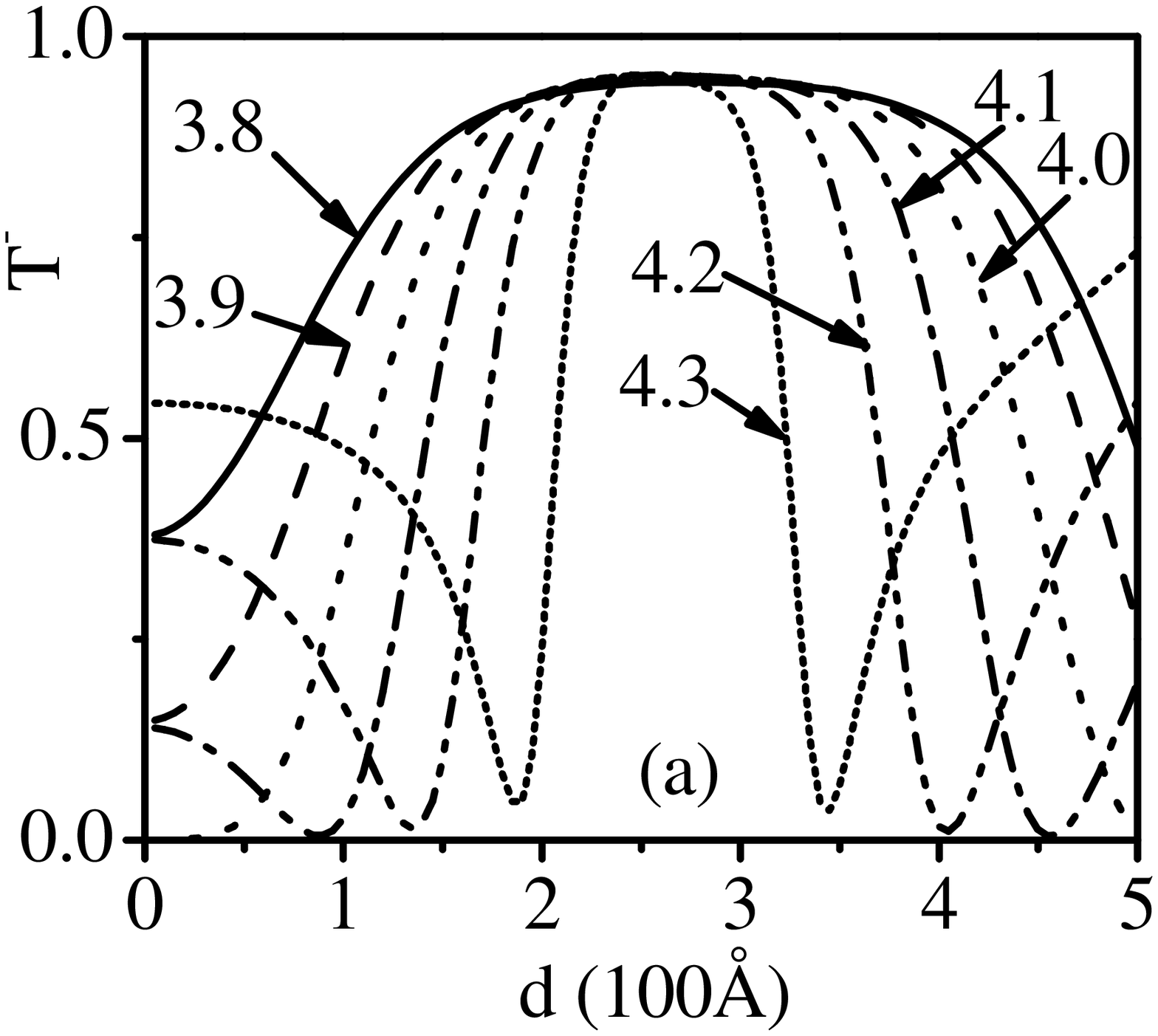}
%\end{figure}
%\begin{figure}[tpb]
%\vspace{-17.15cm}
\includegraphics*[width=70mm,height=70mm]{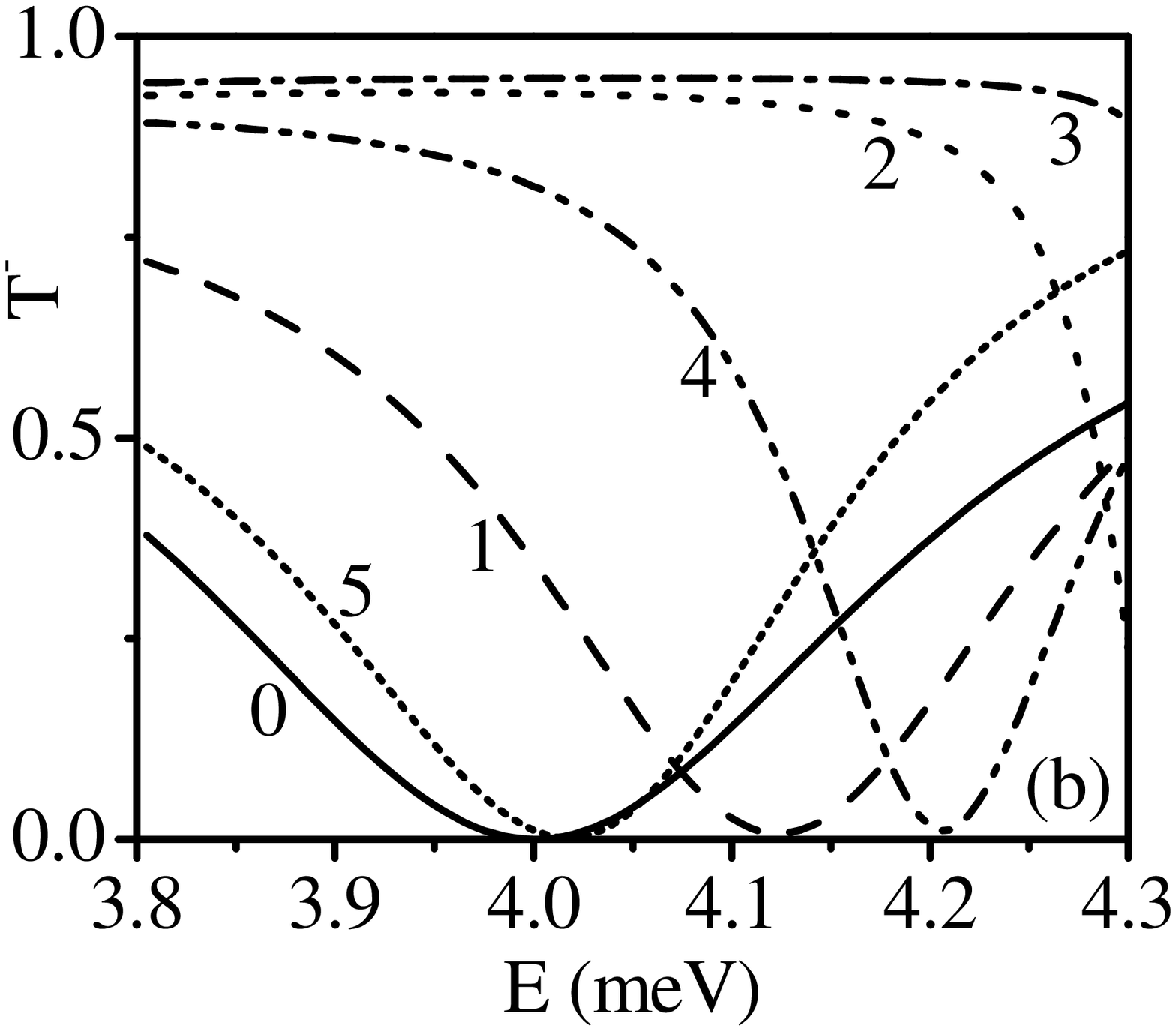}
\vspace{-1.39cm}
\caption{ Spin-down transmission $T^-$, through one stub, as a function of the  asymmetry
parameter $d$ in (a) and of the energy in (b). The various curves are marked by
the value of the energy (in meV) in (a) and by the the value of $d$ (in 100\AA) in (b).}
\label{fig7}
\end{figure}

Now we fix the height $h$ so that all electrons are reflected: for one stub
this happens at $h=1562$\AA\ and for five stubs at
$h=1600$\AA. \ Then we shift the stub along the $x$ direction to change the asymmetry parameter $d$. The result is shown
in Fig. \ref{fig5} and the mixing effect is negligible.  The change in curve
order from (a) to (b) is due the change in the length of the structure.

The transmission through one unit versus the electron energy is shown in Fig. \ref{fig6} for $d=0$.
As in Fig. 3, we observe similar peaks (dips) in the spin-up (spin-down) transmission  due to the subband mixing
and the resulting spin precession. As shown, they occur close to the energies $E=4.36$meV and $E=9.09$meV.
Apart from  these features, the effect  of mixing is negligible.

The effect on the transmission, through one stub,  when we change both the electron energy $E$ and the asymmetry
parameter  $d$, is shown in Fig. \ref{fig7}. In both panels we see the same qualitative behavior between the different
curves: we simply
 notice a shift in the minima (gaps) when the two parameters are varied. If we combine
several stubs  the gaps, e.g., as a function of $d$, become sharper or take a square-wave form as those in Fig. 5.
\begin{figure}[tpb]
\vspace{-0.7cm}
\includegraphics*[width=70mm,height=70mm]{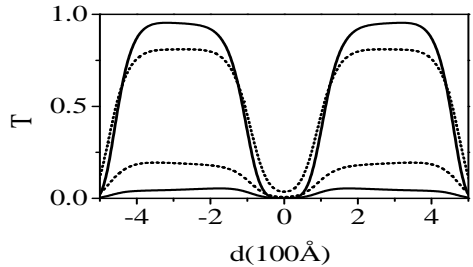}
\vspace{-3.2cm}
\caption{Transmission $T$, through one stub, as a function of the  asymmetry
parameter  $d$  for  $\alpha=1.6\times 10^{-11}$ eVm (solid curves) and $\alpha=1\times 10^{-11}$ eVm (dotted curves).
The upper (lower) curves are for spin down (up).}
\label{fig8}
\end{figure}

Finally, in Fig. 8 we show the dependence of the transmission,  through one stub, on the Rashba parameter $\alpha$. As
can be seen, its qualitative behavior remains the same upon changing $\alpha$. The main change is  in the relative
contributions to the total transmission  of the spin-up  and spin-down states. This results from the phase difference
$k_{y}^{-}-k_{y}^{+}$: when mixing is included, Eq. (4), $k_{y}^{-}-k_{y}^{+}=2m^{\ast }\alpha /\hbar ^{2}$,  is only
approximately satisfied, cf. Fig. 2.

\section{concluding remarks}

We investigated the influence of subband mixing on the 
spin-dependent electronic transmission through periodically stubbed waveguides
in the presence of a weak spin-orbit interaction (SOI). As we  saw, the mixing
affects mainly the  transmission  as a function of the incident electron energy and of the stub height and gives rise to
the prominent peaks (dips) that are absent 
when this mixing is negligible. In contrast, the square-wave pattern of the   transmission  remains robust when the
stubs' degree of asymmetry is varied.  

The results we reported were obtained with parameters more easily accessible to experiments than those of
Ref. {\cite{wan}.
For instance, the waveguide width is twice bigger and the incident (Fermi) energy four times smaller. We hope that this
will further motivate the relevant experiments especially since, as we showed,  a weak mixing leaves almost intact the
square-wave pattern of the transmission  as a function of the stubs' degree of asymmetry.

Though we didn't show any results as a function of the temperature $T$, we verified and can affirm that the $T$
dependence of the  transmission is identical to that reported in Ref. {\cite{wan}. Finite temperatures $T$ smoothen the
curves obtained at $T=0$. As the ratio $E_F/k_BT$
decreases, the curves are smoothened or rounded off more strongly.
        
Finally, we also showed that varying the strength of the SOI parameter changes only the relative contribution to the
total transmission of the spin-up  and spin-down states.  Although   side gates, needed to control the height $h$ and
the distance $d$,  may result in lateral fluctuations, e.g., of $h$, and though they do not directly address the
electron spin, they directly affect the phase of the wave functions in the stubs and accordingly control, through the
matching procedure, the transmission profile of either spin orientation.
As a result, the nearly square-wave pattern of the   transmission  can be made more robust if we combine several units.
This renders the structure we considered  a reasonable candidate for establishing a spin transistor.

{\it Note added in Proof}. The journal reviewers  brought to our attention Ref. \cite{egue}
in which   the effect  of subband mixing,  due to the SOI, on the transmission of electrons with energies near the
anti-crossing energy $E_c$ was considered in one section of  waveguide. The results of this work, obtained in an
approximate way, are similar to ours but apply only to a {\it stubless} waveguide.
%***

\section{Acknowledgements}
 This work
was supported by the  Canadian NSERC Grant No.
OGP0121756.
\clearpage
\section{appendix}
Though not appearing explicitly, the product $\langle\Psi_{nw}^{\sigma}(k)|\Psi_{n'w'}^{\sigma '}(k')\rangle$ in all
matrix products mentioned below denotes
the integral $\int\Psi_{nw}^{\sigma}(k,x)\Psi_{n'w'}^{\sigma '}(k',x)dx$, where $\Psi_{nw}^{\sigma}(x)$ is the
wavefunction along $x$ of a waveguide of width $w$, $n$ is its subband index, $k$  the wavevector, and $\sigma$ the
electron spin. To alleviate the notation
we will denote by $(x_1, x_2,...) {\bf cr}(y_1, y_2,...)$ or $(X) {\bf cr}(Y)$ the product of the {\it column} matrix
 $X$ with the {\it row} matrix $Y$.

The upper ($4\times 8$) part of the $8\times 8$ matrix $\hat{M}_{12}$ is given by
$$
\left(A_h(\gamma)\right) {\bf cr} \left(B_h(\gamma)\right)
\eqno(A1)
$$
where
$$\left(A_h(\gamma)\right)=\left(\langle\Psi_{1h}^{+}(\gamma^+_1)|,\langle\Psi_{1h}^{-}(\gamma^-_1)|,
  \langle\Psi_{2h}^{+}(\gamma^+_2)|,\langle\Psi_{2h}^{-}(\gamma^-_2)|\right)
\eqno(A2)
$$
and 
$$
\left(B_h(\gamma)\right) =\left(|\Psi_{1h}^{+}(\gamma^+_1)\rangle,|\Psi_{1h}^{+}(-\gamma^-_1)\rangle,
   |\Psi_{1h}^{-}(\gamma^-_1)\rangle,|\Psi_{1h}^{-}(-\gamma^+_1)\rangle\right. ,\nonumber \\
$$
$$
\quad\quad  \left. |\Psi_{2h}^{+}(\gamma^+_2)\rangle,|\Psi_{2h}^{+}(-\gamma^-_2)\rangle,
   |\Psi_{2h}^{-}(\gamma^-_2)\rangle,|\Psi_{2h}^{-}(-\gamma^+_2)\rangle\right),
\eqno(A3)
$$
its lower part is
$$\left(A_h(\gamma)\right)  {\bf cr} 
\left(|\Psi_{1h}^{+}(\gamma^+_1)\rangle e^{\gamma^+_1 b},|\Psi_{1h}^{+}(-\gamma^-_1)\rangle e^{-\gamma^-_1 b},
 |\Psi_{1h}^{-}(\gamma^-_1)\rangle e^{\gamma^-_1 b},|\Psi_{1h}^{-}(-\gamma^+_1)\rangle e^{-\gamma^+_1 b}\right. ,\nonumber \\
$$
$$
\quad\quad \left. |\Psi_{2h}^{+}(\gamma^+_2)\rangle e^{\gamma^+_2 b},|\Psi_{2h}^{+}(-\gamma^-_2)\rangle e^{-\gamma^-_2 b},
 |\Psi_{2h}^{-}(\gamma^-_2)\rangle e^{\gamma^-_2 b},|\Psi_{2h}^{-}(-\gamma^+_2)\rangle e^{-\gamma^+_2 b})\right. . 
\eqno(A4)
$$

The lower ($4\times 8$) part of
$\hat{P}_{12}$ ($8\times 8$) is zero; its upper part is given by
$$
\left(A_h(\gamma)\right) {\bf cr} \left(B_c(\eta)\right)
\eqno(A5)
$$

The upper ($4\times 8$) part of
$\hat{Q}_{12}$ ($8\times 8$) is zero; its lower part is the product
$$
\left(A_h(\gamma)\right) {\bf cr} \left(B_a(\beta)\right)
\eqno(A6)
$$

The upper $2\times 4$ part of the matrix $\hat{M}_{n}$ ($n>2$) is given by
$$
\left(C_h(\gamma)\right) {\bf cr}\left(|\Psi_{nh}^{+}(\gamma^+_n)\rangle,|\Psi_{nh}^{+}(-\gamma^-_n)\rangle,
  |\Psi_{nh}^{-}(\gamma^-_n)\rangle,|\Psi_{nh}^{-}(-\gamma^+_n)\rangle\right),
\eqno(A7)
$$
where $$\left(C_h(\gamma)\right)=\left(\langle\Psi_{nh}^{+}(\gamma^+_n)|,\langle\Psi_{1h}^{-}(\gamma^-_n)|\right);
\eqno(A8)
$$  the lower part is 
$$
\left(C_h(\gamma)\right) {\bf cr}\left(|\Psi_{nh}^{+}(\gamma^+_n)\rangle e^{\gamma^+_n b},|\Psi_{nh}^{+}(-\gamma^-_n)\rangle e^{-\gamma^-_n b},
  |\Psi_{nh}^{-}(\gamma^-_n)\rangle e^{\gamma^-_n b},|\Psi_{nh}^{-}(-\gamma^+_n)\rangle e^{-\gamma^+_n b}\right).
\eqno(A9)
$$ 

The lower ($2\times 8$) part of
$\hat{P}_{n}$ ($4\times 8$) is zero; its upper part is given by

$$\left(C_h(\gamma)\right) {\bf cr} \left(B_c(\eta)\right)
\eqno(A10)
$$

The upper ($2\times 8$) part of
$\hat{Q}_{n}$ ($4\times 8$) is  zero; its lower part is the product
$$
\left(C_h(\gamma)\right) {\bf cr} \left(B_a(\beta)\right)
\eqno(A11)
$$

The upper $4\times 8$ part of the $8\times 8$ matrix $\hat{N}_{12}$ is given by
$$ \left(A_c(\eta)\right) {\bf cr}  \left(D_h(\gamma)\right)
\eqno(A12)
$$
where
$$
\left(D_h(\gamma)\right)=\left(\gamma^+_1|\Psi_{1h}^{+}(\gamma^+_1)\rangle,-\gamma^-_1|\Psi_{1h}^{+}(-\gamma^-_1)\rangle,
 \gamma^-_1|\Psi_{1h}^{-}(\gamma^-_1)\rangle,-\gamma^+_1|\Psi_{1h}^{-}(-\gamma^+_1)\rangle\right. ,\nonumber \\
$$
$$
\quad\quad\quad
\ \ \ \ \ \ \ \ \ \ \ \ \ \left. \gamma^+_2|\Psi_{2h}^{+}(\gamma^+_2)\rangle,-\gamma^-_2|\Psi_{2h}^{+}(-\gamma^-_2)\rangle,
 \gamma^-_2|\Psi_{2h}^{-}(\gamma^-_2)\rangle,-\gamma^+_2|\Psi_{2h}^{-}(-\gamma^+_2)\rangle\right) \nonumber
\eqno(A13)
$$
and  the lower one by 
$$ \left(A_a(\beta)\right) {\bf cr}
\left(\gamma^+_1|\Psi_{1h}^{+}(\gamma^+_1)\rangle e^{\gamma^+_1 b},
-\gamma^-_1|\Psi_{1h}^{+}(-\gamma^-_1)\rangle e^{-\gamma^-_1 b},
 \gamma^-_1|\Psi_{1h}^{-}(\gamma^-_1)\rangle e^{\gamma^-_1 b},
-\gamma^+_1|\Psi_{1h}^{-}(-\gamma^+_1)\rangle e^{-\gamma^+_1 b}\right. ,\nonumber \\
$$
$$
\quad\quad\left. \gamma^+_2|\Psi_{2h}^{+}(\gamma^+_2)\rangle e^{\gamma^+_2 b},
-\gamma^-_2|\Psi_{2h}^{+}(-\gamma^-_2)\rangle e^{-\gamma^-_2 b},
 \gamma^-_2|\Psi_{2h}^{-}(\gamma^-_2)\rangle e^{\gamma^-_2 b},
-\gamma^+_2|\Psi_{2h}^{-}(-\gamma^+_2)\rangle e^{-\gamma^+_2 b})\right. .
\eqno(A14)
$$

The lower ($4\times 8$) part
of $\hat{\eta}_{12}$ ($8\times 8$) is zero; its upper part is
the product

$$ \left(A_c(\eta)\right)  {\bf cr} \left(D_c(\eta)\right).
\eqno(A15)
$$ 

The upper ($4\times 8$) part of
$\hat{\beta}_{12}$ ($8\times 8$) is zero; its lower part is given by
$$\left(A_a(\beta)\right)  {\bf cr} \left(D_a(\beta)\right).
\eqno(A16)
$$

The upper $4\times 4$ part of the $8\times 4$ matrix $\hat{N}_{n}$ ($n>2$) is given by

$$ \left(A_c(\eta)\right){\bf cr}
\left(\gamma^+_n|\Psi_{nh}^{+}(\gamma^+_n)\rangle,-\gamma^-_n|\Psi_{nh}^{+}(-\gamma^-_n)\rangle,
   \gamma^-_n|\Psi_{nh}^{-}(\gamma^-_n)\rangle,-\gamma^+_n|\Psi_{nh}^{-}(-\gamma^+_n)\rangle\right),
\eqno(A17)
$$ 
and the lower part by
$$ \left(A_a(\beta)\right){\bf cr}
\left(\gamma^+_n|\Psi_{nh}^{+}(\gamma^+_n)\rangle e^{\gamma^+_n b},
  -\gamma^-_n|\Psi_{nh}^{+}(-\gamma^-_n)\rangle e^{-\gamma^-_n b}\right.,\nonumber\\
$$
$$
\ \ \ \ \ \ \ \left.   \gamma^-_n|\Psi_{nh}^{-}(\gamma^-_n)\rangle e^{\gamma^-_n b},
  -\gamma^+_n|\Psi_{nh}^{-}(-\gamma^+_n)\rangle e^{-\gamma^+_n b}\right). 
\eqno(A18)
$$ 

\end{document}